\documentclass[12pt]{article}
\def \bea{\begin{eqnarray}}
\def \beq{\begin{equation}}
\def \bh{B_H}
\def \bI{\bar I}
\def \bl{B_L}
\def \bo{B^0}
\def \bra#1{\langle #1 |}
\def \bt{\bar t}
\def \ctb{\cos 2 \beta}
\def \da{\downarrow}
\def \dmb{\Delta m \bar t}
\def \dmt{\Delta m t}
\def \eea{\end{eqnarray}}
\def \eeq{\end{equation}}
\def \eb{e^{i \beta}}
\def \emb{e^{- i \beta}}
\def \fdm{\frac{\Delta m t}{2}}
\def \hp{\hat{p}}
\def \ket#1{| #1 \rangle}

\def \Lm{L_{\rm max}}

\def \ob{\overline{B}^0}

\def \pr{\parallel}
\def \s{\sqrt{2}}
\def \stb{\sin 2 \beta}
\def \ua{\uparrow}

\begin{document}

\begin{flushright}
TECHNION-PH-2000-27\\
EFI 2000-33 \\
hep-ph/0010238 \\
October 2000 \\
\end{flushright}

\bigskip
\medskip
\begin{center}
\large
{\bf Tests for Coherence in Neutral ${\bf B}$ Meson Decays
\footnote{To be submitted to Phys.\ Rev.\ D.}}

\bigskip
\medskip

\normalsize
{\it Michael Gronau} \\

\medskip
{\it Department of Physics, Technion-Israel Institute of Technology \\
Technion City, 32000 Haifa, Israel}

\bigskip
and
\bigskip

{\it Jonathan L. Rosner \\
\medskip

Enrico Fermi Institute and Department of Physics \\
University of Chicago, Chicago, Illinois 60637 } \\

\bigskip
\bigskip
{\bf ABSTRACT}

\end{center}

\begin{quote}

A density-matrix method for the study of tagged states of neutral $B$ mesons
with arbitrary coherence properties is applied to several
examples, including $e^+ e^-$ production both at and above the $\Upsilon(4S)$
resonance, and hadronic production.  In the
absence of coherence the only term modulating the exponential decay of a
neutral $B$ meson behaves as $\cos \Delta m t$, while a $\sin \Delta m
t$ modulation is a signal of partial or full coherence.  Decays to CP
eigenstates are needed to fully specify the density matrix.  We relate these
results to more familiar expressions for the cases of the $\Upsilon(4S)$
and incoherent production.

\end{quote}

\section{Introduction}

Neutral $B$ mesons undergo time-dependent oscillations with their
antiparticles.  This feature, first demonstrated
for neutral kaons nearly half a century ago \cite{Kosc}, has been
crucial in extracting fundamental information on the mechanism of CP
violation from the decays of neutral $B$'s.  Moreover, the oscillations
themselves have provided crucial information on the magnitude of
electroweak couplings, and were one of the first pieces of evidence for
a very heavy top quark \cite{heavyt}.

The oscillations are characterized by splittings $\Delta m$ between mass
eigenstates.  For the $B_d = \bar b d$, the most recent world average
\cite{dmd} is $\Delta m_d = 0.487 \pm 0.014$ ps$^{-1}$.  For the $B_s =
\bar b s$, only a lower limit \cite{dmd,dms} $\Delta m_s > 15$ ps$^{-1}$
exists at present.

In the study of CP violation in decays of a neutral $B$ meson, one
frequently needs to know its flavor at the
time of production.  Was it a $\bo$ or a $\ob$?  Was it a $B_s$ or a
$\overline{B}_s$?  The dynamics of $B$ meson production affords several
methods for identifying this flavor.  ``Same-side'' tagging methods
\cite{Ali,btag,dml,dm} utilize the correlation of the flavor of a neutral $B$
with the charge of a kaon or pion which is produced near it in phase
space.  ``Opposite-side'' methods utilize the associated production of
$b \bar b$ in electromagnetic or strong interactions to tag a neutral
$B$ using the fragmentation products of the quark produced in
association with it.  The tagging methods are useful not only for the
study of CP asymmetries, but also in the study of the oscillations
themselves.  For example, it is important to understand the systematic errors
of tagging methods if a reliable estimate of $\Delta m_s$ is to be achieved.

The threshold for electromagnetic or strong production of a pair of
nonstrange $B$ mesons [$M(B) = 5.28$ GeV/$c^2$]
is just below the $\Upsilon(4S)$ resonance, which lies at a center-of-mass
energy of 10.58 GeV.  At the $\Upsilon(4S)$, the reaction $e^+ e^- \to
B \bar B$ produces the two mesons in an eigenstate of the charge
conjugation operactor $C$, with eigenvalue $\eta_C = -1$.  The flavor
oscillations of neutral $B$'s then manifest themselves as functions of the
time {\it difference} $t - \bar t$ between their decays.  Consequently,
asymmetric $e^+ e^-$ collisions have been adopted as a means of
time-dilating the decays to enable the separation of their vertices
\cite{PEP2,KEKB}.

Another means of $B$ production is through the decay $Z \to b \bar b$,
with subsequent fragmentation of the $b$ or $\bar b$ to the neutral
meson of interest.  Here, the observed $B$ and the tagging hadron
(which could be any meson or baryon containing a $b$ or $\bar b$) are
likely to be uncorrelated in their charge-conjugation properties, as
has been assumed in several analyses (e.g., \cite{comb}).

The question of coherence between the tagging hadron and the detected
neutral $B$ meson may not be so clear-cut in several cases intermediate
between $\Upsilon(4S) \to \bo \ob$ (full coherence) and $Z \to b \bar b$
(little or no coherence).  For example, just above
the threshold for $e^+ e^- \to \bo \overline{B}^{*0} {\rm~or~} \ob
B^{*0}$, if the photon in $\overline{B}^{*0} \to \ob \gamma$ or $B^{*0}
\to \bo \gamma$ is detected, the $\bo \ob$ pair will be in a state of
$\eta_C = +1$.  If the photon is not detected, however, there may be an
additional contribution from $e^+ e^- \to \bo \ob$, in which the $\bo
\ob$ pair has $\eta_C = -1$.  The relative probabilities $P_\pm$ of
$\eta_C = \pm 1$ states are in any case unlikely to be equal.

In hadronic $b \bar b$ production the subprocesses $q \bar q \to b \bar
b$ and $g g \to b \bar b$ generate a $b \bar b$ pair whose mass
spectrum peaks at a scale of several times $m_b$.  Additional $b \bar b$
pairs with an even sharper $M(b \bar b)$ peak near threshold arise from
splitting of a virtual gluon: $g^* \to b \bar b$.  While incoherence has
been assumed (e.g., \cite{CDFosc}) in analyses of flavor oscillations in
hadronic $B$ production, there may be effects of coherence if a $\bo$
and a $\ob$ are produced in a state of low enough effective mass.  This
is particularly likely in the case of forward geometries (e.g., the
HERA-B experiment at DESY \cite{HERAB}, the BTeV experiment at Fermilab
\cite{BTeV}, and the LHCb experiment at CERN \cite{LHCb}), in which the
$B$ and $\overline{B}$ are highly kinematically correlated.  It is
less likely to be the case for central $B$ production (as in the CDF and
D0 experiments at Fermilab \cite{CDF,D0} and the ATLAS and CMS experiments
at CERN \cite{ATLAS,CMS}).

The practical effects of coherence cannot be ignored, since they lead to
a characteristic term proportional to $e^{- \Gamma t} \sin \Delta m t$ in
the dependence on proper time $t$ of flavor oscillations.  Such a term
signals unequal probabilities for $\bo \ob$ production in eigenstates of
positive and negative charge-conjugation eigenvalue.  In the absence
of coherence, the only terms present are proportional to $e^{- \Gamma t}$
and $e^{- \Gamma t} \cos \Delta m t$.  We thus advocate the inclusion of
$e^{- \Gamma t} \sin \Delta m t$ terms in any analyses in which the
presence of coherence is suspected.  The present paper is devoted to
the study of such effects.  Although we shall generally speak of the
$(\bo,\ob)$ system, many of our results apply as well to neutral strange
$B$ mesons. 

In Section II we rederive a density-matrix formalism first introduced in
\cite{dml,dm}, using a more standard phase convention and correcting
a sign error in the original references.  This formalism is then applied
to several cases, including mixed states with dilution of tagging
efficiency (Section III), full coherence (Section IV), and intermediate cases
(Section V).  The means of fully specifying the density matrix is discussed in
Section VI.  The degree to which current and planned experiments can be
expected to display coherence is given in Section VII, while Section VIII
concludes.

\section{Density-matrix description}

The density matrix is the appropriate means with which to discuss states
with arbitrary coherence properties.  We work in a two-component
``quasi-spin'' space \cite{BS,HJL} with initial basis states
\beq \label{eqn:fla}
\ket{\bo} = \left[ \begin{array}{c} 1 \\ 0 \\ \end{array} \right]~~~,~~
\ket{\ob} = \left[ \begin{array}{c} 0 \\ 1 \\ \end{array} \right]~~~.
\eeq
In this basis the most general density matrix $\rho$ satisfying $\rho =
\rho^\dag$, Tr$(\rho) = 1$ can be written
\beq
\rho = \frac{1}{2} [ 1 + {\bf Q} \cdot \sigma ]~~~,
\eeq
where ${\bf Q}$ describes polarization in quasispin space, ${\bf Q}^2 \leq 1$,
and $\sigma_i~(i = 1,~2,~3)$ are the Pauli matrices.  A pure state can be
described by a density matrix with $|{\bf Q}| = 1$, while a completely
incoherent combination of $\bo$ and $\ob$ with relative probabilities $P_{\bo}$
and $P_{\ob} = 1 - P_{\bo}$ (a ``mixed state'')
corresponds to a diagonal density matrix with $Q_1 = Q_2
= 0,~Q_3 = 2P_\bo - 1$.  One describes the density matrices for initial $\bo$
and $\ob$ by diag(1,0) and diag(0,1), respectively. 

The probability for a transition from an initial state denoted by the density
matrix $\rho_i$ to a final state denoted by $\rho_f$ is then
\beq
I(f) = {\rm Tr}~(\rho_i {T}^{\dag} \rho_f T)~~~,
\eeq
where $T$ is the operator which time-evolves the state from $i$ to $f$.  Here
$f$ will denote an arbitrary {\it coherent} superposition of $\bo$ and
$\ob$ at time $t$, so that we shall be able to discuss decays to both flavor
eigenstates (such as $J/\psi K^{*0} \to J/\psi K^+ \pi^-$) and CP eigenstates
(such as $J/\psi K_{S,L}$).  The density matrix $\rho_f$ will take the
appropriate form for each such final state.

It is most convenient to transform to the mass eigenstate basis
\beq \label{eqn:mas}
\ket{\bl} = \left[ \begin{array}{c} 1 \\ 0 \\ \end{array} \right]~~~,~~
\ket{\bh} = \left[ \begin{array}{c} 0 \\ 1 \\ \end{array} \right]~~~,
\eeq
where ``L'' denotes ``light'' and ``H'' denotes ``heavy,''
with the relation between mass and flavor eigenstates given by
\beq
\ket{\bl} = p \ket{\bo} + q \ket{\ob}~~,~~~
\ket{\bh} = p \ket{\bo} - q \ket{\ob}~~,~~~
|p|^2 + |q|^2 = 1~~~.
\eeq
In a standard convention \cite{revs} one has $q/p = e^{-2i \beta}$,
where $\beta = {\rm Arg}(- V_{cb}^* V_{cd}/V_{tb}^* V_{td})$, and $V_{ij}$
are elements of the Cabibbo-Kobayashi-Maskawa matrix specifying the
charge-changing weak couplings of quarks.  We then choose $p = \eb/\s,~
q = \emb/\s$.

We shall neglect width differences in the following discussion.  They
are expected to be extremely small for nonstrange neutral $B$'s, although
they may be as large as 10\% for $B_s$ \cite{Bec,Hash}.  In the mass
eigenstate basis (\ref{eqn:mas}) the time evolution operator is
\beq
\bra{M'} T \ket{M} = e^{-i M_D t} = e^{- \Gamma t/2}
{\rm diag}(e^{-i m_L t}, e^{-i m_H t})
= e^{- \Gamma t/2} e^{- i \bar m t} e^{i \sigma_3 \Delta m t /2}~~~,
\eeq
where $\bar m \equiv (m_H + m_L)/2$ and $\Delta m \equiv m_H - m_L$.
Transforming to the flavor basis (\ref{eqn:fla}), we find
\beq
\bra{F'} T \ket{F} = e^{- \Gamma t/2} e^{- i \bar m t} \left[
\begin{array}{c c} \cos \fdm & i e^{2i \beta} \sin \fdm \\
i e^{- 2 i \beta} \sin \fdm & \cos \fdm \end{array} \right]~~~.
\eeq

It is most convenient to express the density matrix in the mass basis as
well:
\beq
\bra{M'} \rho \ket{M} = \sum_{F,F'} \bra{M'} F' \rangle \bra{F'} \rho
\ket{F} \langle F \ket{M}~~~.
\eeq
We shall denote the density matrix in the mass basis by
\beq
\rho' = \frac{1}{2} \left[ 1 + {\bf Q'} \cdot \sigma \right]~~~;
\eeq
the vector $Q'$ in the mass basis is related to the vector $Q$ in the
flavor basis by
\bea
Q_1' & = & Q_3~~~, \\
Q_2' & = & -(Q_1 \stb + Q_2 \ctb)~~~, \\
Q_3' & = & Q_1 \ctb - Q_2 \stb~~~.
\eea
Since states which are pure $\bo$ or pure $\ob$ correspond to $Q_3 = \pm 1,~
Q_1 = Q_2 = 0$, their transformed density matrices are $\rho'_f = (1/2)
(1 \pm \sigma_1)$, respectively, since then $Q_1' = \pm 1,~ Q_2' = Q_3' = 0$.
More generally, an incoherent state with $Q_3 = 2 P_{\bo} - 1$, $Q_1 =
Q_2 = 0$ corresponds to $\rho'_f = (1/2) [1 + (2 P_{\bo} - 1) \sigma_1]$.

The transition probability can now be written in terms of traces as 
\beq \label{treq}
I(f) = {\rm Tr}~ (\rho_i' e^{i M^*_D t} \rho_f' e^{-i M_D t} )~~~.
\eeq
For flavor eigenstates $f$ corresponding to $\bo$ or $\ob$, let us take
as examples $J/\psi K^{*0} \to J/\psi K^+ \pi^-$ and $J/\psi
\overline{K}^{*0} \to J/\psi K^- \pi^+$, respectively.  In the present
convention both decay amplitudes are equal to the same constant $A$,
since they involve the quark subprocesses $\bar b \to \bar c c \bar s$
and $b \to c \bar c s$, respectively.  Then we find
\beq \label{eqn:int}
I \left( \begin{array}{c} \bo \\ \ob \end{array} \right) \propto
e^{- \Gamma t} [ 1 \pm (Q_1' \cos \dmt + Q_2' \sin \dmt) ]~~~.
\eeq
The sign in front of the $Q_2'$ term was incorrectly stated in Refs.\
\cite{dml} and \cite{dm}. 

As noted in Refs.\ \cite{dml} and \cite{dm}, the component $Q_3'$ does
not appear in these expressions.  We shall return to the question of its
determination in Section VI.

\section{Mixed state}

A mixed state of $\bo$ and $\ob$ is one in which there are no amplitude
correlations between the $\bo$ and $\ob$.  Such a state will arise, in
general, when a $b \bar b$ pair is produced with high enough effective mass
that the $b$ and $\bar b$ fragment independently.  In this case we can
consider a tagging method to indicate with probability $P_r$ the right-sign
neutral $B$ and with probability $P_w = 1 - P_r$ the wrong-sign neutral $B$.
The {\it dilution factor} ${\cal D}$ is ${\cal D} = P_r - P_w = 2 P_r - 1$.

Dilution can occur in various ways, depending on the tagging method.  In
opposite-side tagging at high $M(b \bar b)$, the opposite-side quark may
fragment into a charged or neutral nonstrange $B$ meson, a strange $B$
meson, or a beauty baryon.  These fractions have been measured at CDF
\cite{CDFfr} and LEP \cite{dmd} and are summarized in Table I.

\begin{table}
\caption{Fractions $f(H)$ of hadrons produced in $b$ quark fragmentation.}
\begin{center}
\begin{tabular}{c c c} \hline
Hadron & CDF (a) & LEP \\ \hline
$\ob$ & $0.375 \pm 0.023$ & $0.40 \pm 0.01$ \\
$B^-$ & $0.375 \pm 0.023$ & $0.40 \pm 0.01$ \\
$\overline{B}_s$ & $0.160 \pm 0.044$ & $0.097 \pm 0.012$ \\
Baryons & $0.090 \pm 0.029$ & $0.104 \pm 0.017$ \\ \hline
\end{tabular}
\end{center}
\leftline{(a) Assuming equal fractions of charged and neutral nonstrange $B$
mesons.}
\end{table}

The probability that a $\ob$
is detected as a $\bo$ is $x_d^2/[2(1+x_d^2)] \simeq 0.18$, where $x_d \equiv
\Delta m_d/\Gamma_d = (0.487 \pm 0.014~{\rm ps}^{-1})(1.56 \pm 0.04~{\rm
ps}) = 0.76 \pm 0.03$ \cite{dmd,PDG}.  The corresponding probability of
mis-detecting the flavor of a $\overline{B}_s$ is very close to 1/2.
Assuming that the other flavors are detected with unit probability, the
CDF results imply $P_r \simeq 0.85$ and ${\cal D} = 0.70$
while the LEP results imply $P_r \simeq 0.88$ and ${\cal D} = 0.76$.  In
practice many other factors of course contribute to the dilution of a
tagging method.

Given a tag which should indicate the presence of a $\bo$ at time of
production with tagging probability $P_{\bo}$, the corresponding density
matrix elements in the mass basis are $Q_1' = 2 P_{\bo} - 1$, $Q_2' = Q_3' =
0$.  The time-dependence of the flavor-specific final state $f$ arising
from either a $\bo$ or $\ob$ decay is then
given by
\beq
I \left( \begin{array}{c} \bo \\ \ob \end{array} \right) \propto
e^{- \Gamma t} [ 1 \pm (2 P_{\bo} - 1) \cos \dmt ]~~~,
\eeq
without any $\sin \dmt$ term.  The quantity $P_{\bo}$ is usually determined
empirically in a fit which also yields $\Delta m$.

\section{Full coherence}

We now consider the case of fully coherent states of $\bo$ and $\ob$ produced
in states of definite charge conjugation eigenvalue.  We denote a $C$
eigenstate of $\bo$ and $\ob$ by
\beq
\Psi_C = \frac{1}{\s} \left[ \bo(\hp) \ob(-\hp) + \eta_C \ob(\hp)
\bo(-\hp) \right]~~~,
\eeq
where $\eta_C = \pm 1$ is the eigenvalue of the $C$ operator, and $\hp$
and $-\hp$ are unit vectors denoting the direction of the particles in
their center-of-mass.  The case of $\Upsilon(4S) \to \bo \ob$
corresponds to $\eta_C = -1$.

The states $\Psi_C$ can be written in terms of mass eigenstates as
\bea
\Psi_C(\eta_C = -1) & = & \frac{1}{\s} \left[ B_H(\hp) B_L(-\hp) -
B_L(\hp) B_H(-\hp) \right]~~~, \\
\Psi_C(\eta_C = +1) & = & \frac{1}{\s} \left[ B_L(\hp) B_L(-\hp) -
B_H(\hp) B_H(-\hp) \right]~~~.
\eea
These expressions allow us to write the elements $Q_i'$ of the density
matrix in the mass-eigenstate representation and thereby to calculate
the correlations between particles traveling along $\hp$ (decaying at
proper time $t$) and those traveling along $-\hp$ (decaying at proper
time $\bt$).  We shall derive the results using both the one-particle
formalism given in Section II and a two-particle formalism more suitable
for joint distributions.
\bigskip

\leftline{\bf A.  One-particle description}
\bigskip

We consider for definiteness the case in which a flavor tag ($\bo$ or
$\ob$) is applied at a time $\bt$.  Recalling that
\beq
\langle \bo \ket{B_L(-\hp)} = \langle \bo \ket{B_H(-\hp)} =
\frac{1}{\s}\eb~~,~~~
\langle \ob \ket{B_L(-\hp)} = - \langle \ob \ket{B_H(-\hp)} =
\frac{1}{\s}\emb~~~,
\eeq
we can calculate the dependence on the tagging particle's decay time $\bt$
of the production amplitudes of the states $\ket{B_{L,H}(\hp)}$.
For example, the state $\ket{B_L(\hp)}$ appears in (17) with
coefficient $-\ket{B_H(-\hp)}/\s$, so it depends on $\bt$ with coefficient
$-(1/2)e^{i \beta} e^{-\Gamma \bt/2}e^{-i m_H \bt}$ for a $\bo$ tag and
$(1/2)e^{-i \beta} e^{-\Gamma \bt/2}e^{-i m_H \bt}$ for a $\ob$
tag.  For states with both values of $\eta_C$ we then find
\bea
\eta_C & = & -1, \bo~{\rm tag}:~~~ \left[ \begin{array}{c}
\ket{B_L(\hp)} \\ \ket{B_H(\hp)} \end{array} \right] =
\frac{1}{2} \eb e^{-\Gamma \bt/2}
\left[ \begin{array}{c} -e^{- i m_H \bt} \\ e^{-i m_L \bt} \end{array}
\right]~~~, \\
\eta_C & = & -1, \ob~{\rm tag}:~~~ \left[ \begin{array}{c}
\ket{B_L(\hp)} \\ \ket{B_H(\hp)} \end{array} \right] =
\frac{1}{2} \emb e^{-\Gamma \bt/2}
\left[ \begin{array}{c} e^{- i m_H \bt} \\ e^{-i m_L \bt} \end{array}
\right]~~~, \\
\eta_C & = & +1, \bo~{\rm tag}:~~~ \left[ \begin{array}{c}
\ket{B_L(\hp)} \\ \ket{B_H(\hp)} \end{array} \right] =
\frac{1}{2} \eb e^{-\Gamma \bt/2}
\left[ \begin{array}{c} e^{-i m_L \bt} \\ -e^{-i m_H \bt} \end{array}
\right]~~~, \\
\eta_C & = & +1, \ob~{\rm tag}:~~~ \left[ \begin{array}{c}
\ket{B_L(\hp)} \\ \ket{B_H(\hp)} \end{array} \right] =
\frac{1}{2} \emb e^{-\Gamma \bt/2}
\left[ \begin{array}{c} e^{-i m_L \bt} \\ e^{-i m_H \bt} \end{array}
\right]~~~.
\eea
Translating these pure states into normalized density matrices with
unit trace, we find the results summarized in Table II.  The component
$Q'_3$ is zero.

\begin{table}
\caption{Density matrix elements corresponding to correlated $\bo \ob$
production in states of definite charge-conjugation eigenvalue
$\eta_C$.}
\begin{center}
\begin{tabular}{r c c c} \hline
$\eta_C$ & Tag ($-\hp$) &    $Q_1'$    &    $Q_2'$    \\ \hline
$-1$     &    $\bo$     & $-\cos \dmb$ & $-\sin \dmb$ \\
         &    $\ob$     & $ \cos \dmb$ & $ \sin \dmb$ \\
$+1$     &    $\bo$     & $-\cos \dmb$ & $ \sin \dmb$ \\
         &    $\ob$     & $ \cos \dmb$ & $-\sin \dmb$ \\ \hline
\end{tabular}
\end{center}
\end{table}

The density matrix elements in Table II can now be combined with the
expression (\ref{eqn:int}) to give joint rates for production of states with
direction $\hp$ decaying at time $t$ and direction $-\hp$ decaying at
time $\bt$ \cite{fact}.  We find:
\bea \label{eqn:jt}
I[\bo(t),\bo(\bt)] = I[\ob(t),\ob(\bt)] & = & e^{-\Gamma(t+\bt)} [1 - \cos
\Delta m (t + \eta_C \bt)]~~~, \\
I[\bo(t),\ob(\bt)] = I[\ob(t),\bo(\bt)] & = & e^{-\Gamma(t+\bt)} [1 + \cos
\Delta m (t + \eta_C \bt)]~~~.
\eea

The above expressions are consistent with those in the literature (e.g.,
\cite{Bigrev}) and make physical sense.  Their dependence on $t + \eta_C
\bt$ is mandated by Bose statistics.  When $\eta_C = -1$ and $t = \bt$,
one never sees the decay products of neutral $B$ mesons of the same
flavor.  This is also the case for $t=\bt=0$ when $\eta_C = +1$. 
When tags of each flavor are combined, the oscillatory terms cancel one
another and one is left with a pure exponential $\sim e^{- \Gamma(t + \bt)}$.
\bigskip

\leftline{\bf B.  Two-particle description}
\bigskip

For entangled states such as described by $\Psi_C$ in Eqs.~(17) and (18),
one can use a direct-product notation \cite{Mann}.  Our convention will be
such that the first state in the direct product refers to the particle with
direction $\hp$, while the second refers to that with $-\hp$.  Typical
direct products are then
\beq
\left[ \begin{array}{c} 1 \\ 0 \end{array} \right] \otimes
\left[ \begin{array}{c} 0 \\ 1 \end{array} \right] =
\left[ \begin{array}{c} 0 \\ 0 \\ 1 \\ 0 \end{array} \right]~~,~~~
\left[ \begin{array}{c} 0 \\ 1 \end{array} \right] \otimes
\left[ \begin{array}{c} 1 \\ 0 \end{array} \right] = 
\left[ \begin{array}{c} 0 \\ 1 \\ 0 \\ 0 \end{array} \right]~~~.
\eeq
A spin-singlet state of two spin-1/2 particles is then represented by
the four-component vector
\beq
\ket{S = 0} = \frac{1}{\s}(\ua \da - \da \ua) = \frac{1}{\s} \left[
\begin{array}{r} 0 \\ -1 \\ 1 \\ 0 \end{array} \right]
\eeq
and by the density matrix
\beq
\rho(S=0) = \frac{1}{2} \left[ \begin{array}{r r r r} 0 & 0 & 0 & 0 \\
0 & 1 & -1 & 0 \\ 0 & -1 & 1 & 0 \\ 0 & 0 & 0 & 0 \end{array} \right]~~~.
\eeq
In terms of the direct product representation, this can be written as
\beq
\rho(S=0) = \frac{1}{4} I \otimes I - \frac{1}{4} \sigma_1 \otimes \sigma_1
- \frac{1}{4} \sigma_2 \otimes \sigma_2 - \frac{1}{4} \sigma_3 \otimes
\sigma_3~~~,
\eeq
where $I$ is the $2 \times 2$ identity matrix.  We have used the identities
$$
\sigma_1 \otimes \sigma_1 = \left[ \begin{array}{r r r r} 0 & 0 & 0 & 1 \\
0 & 0 & 1 & 0 \\ 0 & 1 & 0 & 0 \\ 1 & 0 & 0 & 0 \end{array} \right]~~,~~~
\sigma_2 \otimes \sigma_2 = \left[ \begin{array}{r r r r} 0 & 0 & 0 & -1 \\
0 & 0 & 1 & 0 \\ 0 & 1 & 0 & 0 \\ -1 & 0 & 0 & 0 \end{array} \right]~~~,
$$
\beq
\sigma_3 \otimes \sigma_3 = \left[ \begin{array}{r r r r} 1 & 0 & 0 & 0 \\
0 & -1 & 0 & 0 \\ 0 & 0 & -1 & 0 \\ 0 & 0 & 0 & 1 \end{array} \right]~~,~~~
I \otimes I = \left[ \begin{array}{r r r r} 1 & 0 & 0 & 0 \\
0 & 1  & 0 & 0 \\ 0 & 0 & 1 & 0 \\ 0 & 0 & 0 & 1 \end{array} \right]~~~.
\eeq
Other useful identities are
\beq
\sigma_1 \otimes \sigma_2 = \left[ \begin{array}{r r r r} 0 & 0 & 0 & -i \\
0 & 0 & -i & 0 \\ 0 & i & 0 & 0 \\ i & 0 & 0 & 0 \\\end{array} \right]~~,~~~
\sigma_2 \otimes \sigma_1 = \left[ \begin{array}{r r r r} 0 & 0 & 0 & -i \\
0 & 0 & i & 0 \\0 & -i & 0 & 0 \\ i & 0 & 0 & 0 \\\end{array} \right]~~~.
\eeq
Writing the states and their time-evolution as in the previous subsection,
we then find the corresponding density matrices
$$
\rho(\eta_C = -1) = \frac{1}{2} e^{- \Gamma(t + \bt)} \left[ \begin{array}
{c c c c} 0 & 0 & 0 & 0 \\ 0 & 1 & -e^{-i \Delta m (t - \bt)} & 0 \\
0 & -e^{i \Delta m (t - \bt)} & 1 & 0 \\ 0 & 0 & 0 & 0 \end{array} \right]
= \frac{1}{4}e^{-\Gamma(t+\bt)} \left[ I \otimes I - \sigma_3 \otimes \sigma_3
\right.
$$
\beq \left.
-(\sigma_1 \otimes \sigma_1 + \sigma_2 \otimes \sigma_2)\cos \Delta m(t-\bt)
-(\sigma_1 \otimes \sigma_2 - \sigma_2 \otimes \sigma_1)\sin \Delta m(t-\bt)
\right]~~~,
\eeq
$$
\rho(\eta_C = +1) = \frac{1}{2} e^{- \Gamma(t+\bt)} \left[ \begin{array}
{c c c c} 1 & 0 & 0 & -e^{i \Delta m (t + \bt)} \\ 0 & 0 & 0 & 0
\\ 0 & 0 & 0 & 0 \\ -e^{- i \Delta m (t + \bt)} & 0 & 0 & 1 \end{array}
\right] = \frac{1}{4}e^{-\Gamma(t+\bt)} \left[ I \otimes I + \sigma_3
\otimes \sigma_3 \right.
$$
\beq \left.
- (\sigma_1 \otimes \sigma_1 - \sigma_2 \otimes \sigma_2)
\cos \Delta m (t + \bt) + (\sigma_1 \otimes \sigma_2 + \sigma_2 \otimes 
\sigma_1) \sin \Delta m (t + \bt) \right]~~~.
\eeq
These results may be used to derive such expressions as (24) and (25) in
an alternate way.

\section{Intermediate cases}

The density matrices for $\eta_C = +1$ and $\eta_C = -1$ can be added to
one another.  If the probability of states with $\eta_C = \pm 1$ is
denoted by $P_\pm$ with $P_+ + P_- = 1$, the resulting elements for
$\bo(\hp)$ production with a $\bo(-\hp)$ tag are
\beq
\bo(\hp),\bo(-\hp):~~~Q'_1 = -\cos \dmb~~,~~~
Q'_2 = (P_+ - P_-) \sin \dmb~~~,
\eeq
with the signs of $Q'_{1,2}$ changed for a $\ob(-\hp)$ tag.
The joint probabilities for production of (opposite,same) flavors of
neutral $B$ mesons at times $t$ and $\bt$ are then
\beq \label{eqn:ttb}
I \left( \begin{array}{c} {\rm Opp} \\ {\rm Same} \end{array} \right)(t,\bt) =
e^{- \Gamma(t + \bt)} [1 \pm \cos \dmt \cos \dmb \pm (P_- - P_+) 
\sin \dmt \sin \dmb]~~~.
\eeq
  
Equation (\ref{eqn:ttb}) can be integrated with respect to time, with
the result
\beq
\int_0^\infty d \bt I \left( \begin{array}{c} {\rm Opp} \\ {\rm Same}
\end{array} \right)(t,\bt) = \frac{1}{\Gamma} e^{- \Gamma t} \left[
1 \pm \frac{1}{1+ x^2} \cos \dmt \pm (P_- - P_+) \frac{x}{1 + x^2}
\sin \dmt \right]~~~,
\eeq
where $x \equiv \Delta m/\Gamma$.  As long as $P_- \ne P_+$, the $\sin \dmt$
term will be present.

The two-particle description can also be applied to intermediate cases.
For a state which is a mixture of $\eta_C = + 1$ with probability $P_+$ and
$\eta_C = -1$ with probability $P_- = 1 - P_+$, the density matrix is
$$
\rho = \frac{1}{2}e^{-\Gamma(t+\bt)} \left[ \begin{array}{c c c c}
P_+ & 0 & 0 & -P_+ e^{i \Delta m (t + \bt)} \\
0 & P_- & -P_- e^{-i \Delta m (t - \bt)} & 0 \\
0 & -P_- e^{i \Delta m (t - \bt)} & P_- & 0 \\
- P_+ e^{-i \Delta m (t + \bt)} & 0 & 0 & P_+ \end{array} \right]
$$
$$
= \frac{1}{4}e^{-\Gamma(t+\bt)} \left\{ I \otimes I + (P_+ - P_-) \sigma_3
\otimes \sigma_3 - \left[ \begin{array}{c c} 0 & e^{i \dmt} \\ e^{- i \dmt} &
0 \end{array} \right] \otimes \left[ \begin{array}{c c} 0 & e^{i \dmb} \\
e^{-i \dmb} & 0 \end{array} \right] \right.
$$
\beq \left.
+(P_- - P_+) \left[ \begin{array}{c c} 0 & e^{i \dmt} \\ -e^{- i \dmt} & 0
\end{array} \right] \otimes \left[ \begin{array}{c c} 0 & e^{i \dmb} \\
-e^{- i \dmb} & 0 \end{array} \right] \right\}~~~.
\eeq 
This last equation says, in particular, that in order to specify a one-particle
state in which the matrix element $Q'_3$ is non-zero, one must not only
have $P_+ \ne P_-$, but the ``tagging'' particle (with decay time $\bt$)
must also correspond to non-zero $Q'_3$.

These expressions hold for both nonstrange and strange neutral $B$
mesons.  They must be modified to take account of dilution effects such
as those discussed in Section III. However, such effects should reduce the
coefficients of the $\cos \dmt$ and $\sin \dmt$ terms by a common factor.

Since $x$ is very large for $B_s$ mesons, the presence of the $\sin
\dmt$ term may be difficult to demonstrate for them, unless one resolves
the dependence on the ``tagging'' time $\bt$ and does not integrate with
respect to it.

\section{Full specification of the density matrix}

As pointed out in Refs.\ \cite{dml,dm}, it is necessary to observe
decays to CP eigenstates and not just to flavor eigenstates in order to
fully specify the density matrix, since the element $Q'_3$ does not
appear in any of the previous expressions for rates.  We consider decays
to $J/\psi K_S$ and $J/\psi K_L$.

Taking account of the negative CP of $J/\psi K_S$ and positive CP of
$J/\psi K_L$, the decay amplitudes of interest are
\beq
\bra{J/\psi K_S} \bo \rangle = - \bra{J/\psi K_S} \ob \rangle =
\bra{J/\psi K_L} \bo \rangle = \bra{J/\psi K_L} \ob \rangle = A'/\s~~~.
\eeq 
Then the density matrices for each final state in the flavor basis are 
\beq
\rho_{J/\psi K_S} \frac{1}{2}|A'|^2 \left[ \begin{array}{r r} 1 & -1 \\
-1 & 1 \end{array} \right]~~,~~~\rho_{J/\psi K_L} = \frac{1}{2}|A'|^2 \left[
\begin{array}{r r} 1 & 1 \\ 1 & 1 \end{array} \right]~~~,
\eeq
while in the mass-eigenstate basis they are
\beq
\rho'_{J/\psi K_S} = \frac{1}{2}|A'|^2 \left[ \begin{array}{c c}
1 - \ctb & -i \stb \\ i \stb & 1 + \ctb \end{array} \right]~~~,
\eeq
\beq
\rho'_{J/\psi K_L} = \frac{1}{2}|A'|^2 \left[ \begin{array}{c c}
1 + \ctb & i \stb \\ -i \stb & 1 - \ctb \end{array} \right]~~~.
\eeq
We then recover the results of Refs.\ \cite{dml,dm}, aside from a sign
in the $Q'_2$ term which we correct here:
\bea
I \left( J/\psi \begin{array}{c} K_L \\ K_S \end{array} \right) & \propto &
e^{-\Gamma t} \left\{ 1 \pm [Q'_3 \ctb + (Q'_1
\sin \dmt - Q'_2 \cos \dmt) \stb \right\}, \nonumber \\
\bI \left( J/\psi \begin{array}{c}K_L \\ K_S \end{array} \right)
& \propto & e^{-\Gamma t} \left\{ 1 \pm [Q'_3 \ctb - (Q'_1
\sin \dmt - Q'_2 \cos \dmt) \stb \right\}, \nonumber
\end{eqnarray}
where $I$ refers to a rate tagged with an opposite-side $B$, while $\bar I$
refers to a rate tagged with an opposite-side $\overline{B}$.

The determinations of $Q'_3$ and $\ctb$ are interrelated.  Information on the
sign of $\ctb$ would be useful in resolving the discrete ambiguity associated
with extracting the value of $\beta$ from that of $\stb$ \cite{KS}.  However,
in eigenstates of $C$ with $\eta_C = \pm 1$, the two-particle density matrix
results indicate that the contributions from $\sigma_3$ for the particles
decaying at times $t$ and $\bt$ are correlated.  Thus, in order to prepare a
state with $Q'_3 \ne 0$ decaying at time $t$ it appears that one must {\it tag}
with a CP eigenstate decaying at time $\bt$.  For example, in
$e^+ e^- \to \Upsilon(4S) \to \bo \ob \to (J/\psi K_{S,L})(\hp)(J/\psi K_{S,L})
(-\hp)$, the effects of $Q'_3 \ne 0$ will always involve the term $\cos^2 2
\beta$, so information on the sign of $\ctb$ is lost.

An explicit calculation with the two-particle density matrix leads to the
following time-dependent rates for a mixture of C eigenstates with
probabilities $P_+$ and $P_- = 1 - P_+$:
$$
\frac{d^2 \Gamma}{d t~d \bt}[J/\psi K_{S,L}(t)J/\psi K_{S,L}(\bt)]
\propto \{1 + (P_+ - P_-)\cos^2 2 \beta
$$
\beq
- \sin^2 2 \beta [ \sin \dmt \sin \dmb + (P_- - P_+) \cos \dmt \dmb]
\}~~~,
\eeq
$$
\frac{d^2 \Gamma}{d t~d \bt}[J/\psi K_{S,L}(t)J/\psi K_{L,S}(\bt)]
\propto \{1 - (P_+ - P_-)\cos^2 2 \beta
$$
\beq
+ \sin^2 2 \beta [ \sin \dmt \sin \dmb + (P_- - P_+) \cos \dmt \dmb]
\}~~~.
\eeq
As noted, $\beta$ appears only through $\sin^2 2 \beta$ and $\cos^2 2 \beta
= 1 - \sin^2 2 \beta$.

In principle the reaction $e^+ e^- \to \Upsilon(4S) \to \bo \ob \to
(J/\psi K_{S,L})(\hp) (\pi^+ \pi^-)(-\hp)$ can provide additional information.
The (normalized) density matrices for $\pi^+ \pi^-$ production are
\beq
\rho_{\pi^+ \pi^-} = \frac{1}{2} \left[ \begin{array}{c c} 1 & e^{-i \gamma} \\
e^{i \gamma} & 1 \end{array} \right]
\eeq
in the flavor basis and
\beq
\rho'_{\pi^+ \pi^-} = \frac{1}{2} \left[ \begin{array}{c c} 1 + \cos 2 \alpha &
- i \sin 2 \alpha \\ i \sin 2 \alpha & 1 - \cos 2 \alpha \end{array} \right]
\eeq
in the mass-eigenstate basis.  (We have neglected the effect of penguin
amplitudes here.)  The use of Eq.~(37) then leads to expressions involving
the combinations $\cos 2 \beta~\cos 2 \alpha$ and $\sin 2 \beta~\sin 2 \alpha$. 
 
In a state of definite charge-conjugation eigenvalue we find, by
substituting the values in Table II and noting that $Q'_3 =0$ for such
a state, that
\bea
I \left( J/\psi \begin{array}{c} K_L \\ K_S \end{array} \right) &
\propto & e^{-\Gamma(t + \bt)} \left[ 1 \mp \stb \sin \Delta
m (t + \eta_C \bt) \right]~~~, \\
\bI \left( J/\psi \begin{array}{c}K_L \\ K_S \end{array} \right) &
\propto & e^{-\Gamma(t + \bt)} \left[ 1 \pm \stb \sin \Delta
m (t + \eta_C \bt) \right]~~~.
\eea
The results for $\eta_C = -1$ agree with those in Ref.\ \cite{Bigrev},
while for $\eta_C = +1$ the sign of the $\stb \sin \Delta m (t + \bt)$
term is reversed.  Again, although we have used the one-particle expressions
based on the vector ${\bf Q'}$, these results can also be derived using
the two-particle density matrices (32) and (33).

\section{Coherence expected in present and planned experiments}

The specific cases we have discussed so far range from fully coherent
$\bo \ob$ production at the $\Upsilon(4S)$ to uncorrelated production
at high effective $b \bar b$ masses (as in $Z^0 \to b \bar b$).  A
qualitative estimate of the degree of coherence expected between
$\bo$ and $\ob$ may be obtained by examining their relative orbital
angular momenta.  Since for a meson-antimeson pair with relative orbital
angular momentum $L$ the charge-conjugation eigenvalue is $\eta_C =
(-1)^L$, the degree of coherence is expected to decrease as the
accessible values of $L$ increase.

Suppose, for example, that the population of orbital angular momentum
levels of $\bo \ob$ is dictated by their statistical weights $2 L + 1$
up to a maximum $L = \Lm$.  If $\Lm=0$ the probability $P_+$ of $\eta_C
= 1$ is 1, while the probability $P_-$ of $\eta_C = -1$ is 0.  If
$\Lm=1$ then $P_+ = 1/4$, $P_- = 3/4$.  The general expressions are
\bea
\Lm~{\rm even}:~~~P_+ & = & \frac{\Lm + 2}{2(\Lm+1)}~~,~~~
P_- = \frac{\Lm}{2(\Lm+1)}~~~,\\
\Lm~{\rm odd}:~~~~P_+ & = & \frac{\Lm}{2(\Lm+1)}~~,~~~
P_- = \frac{\Lm + 2}{2(\Lm+1)}~~~.
\eea
Then one finds $P_+ - P_- = (-1)^{\Lm}/(\Lm+1)$, and the magnitude of
the coefficient of the $\sin \dmt$ term in the time-dependence of a
flavor eigenstate with a flavor tag decreases as $1/\Lm$.

A semiclassical argument can be used to estimate $\Lm$.  Imagine a $b
\bar b$ pair with squared c.m.\ energy $s$ to fragment into a pair of $B$
mesons.  The fragmentation process is limited to impact parameters $b_0
\le 1$ fm $\simeq 5$ GeV$^{-1}$.  Thus
\beq
\Lm = k b_0~~,~~~k \equiv \sqrt{(s/4) - m_b^2}~~~.
\eeq
We now discuss the specific experimental cases mentioned in the
Introduction, in decreasing order of likelihood of $\bo \ob$ coherence.

(1) {\it Production at the $\Upsilon(4S)$} leads to a $\bo \ob$ pair in
a state with $L = 1$, $\eta_C = -1$.  The full-coherence arguments of
Section III apply.

(2) {\it Production through the process $e^+ e^- \to \bo \overline{B}^{*0}
+ {\rm c.c.}$} gives rise to a $\bo \ob$ pair with $\eta_C = +1$, since
$B^{*0}$ decays entirely to $\gamma \bo$.

(3) {\it Contamination of $e^+ e^- \to \bo \overline{B}^{*0} + {\rm c.c.}$
by $e^+ e^- \to \bo \ob$} is likely.  At the threshold for (vector +
pseudoscalar) meson production, pseudoscalar meson pair production is also
kinematically allowed, leading to some admixture of the $\eta_C = -1$ state.
The relative fractions of the two processes are not well known
but can be measured in principle.  One can ensure against the $\eta_C =
-1$ state by detecting the 46 MeV photon in $B^{*0}$ decay.

(4) {\it Forward hadronic production of $\bo \ob$} will lead to a pair with
effective mass typically no more that a few times the threshold mass
of $2 m_b$.  The subprocesses $q \bar q \to b \bar b$ and $g g \to b
\bar b$ both favor low effective $b \bar b$ masses, while the
gluon-splitting process $g^* \to b \bar b$ will favor even lower
effective masses.  The corresponding value of $k$ will then be of order $m_b$,
leading to $\Lm = {\cal O}[m_b \cdot (1 {\rm~fm})] \simeq 25$.  Thus
one might expect magnitudes of $P_- - P_+$ of at most a few percent.

(5) {\it Central hadronic production} may lead to somewhat higher
effective $b \bar b$ masses, especially if ``opposite-side'' tagging
utilizes $B$'s produced in the opposite hemisphere of the detector.  The
probability of $\bo \ob$ coherence is thus likely to be less than in
forward geometries.

(6) {\it Production in $Z \to b \bar b$} is expected to lead to very
little $\bo \ob$ coherence, since $k \simeq M_Z/2$ and $\Lm$ consequently
exceeds 200.

The best prospect for studying the coherence effects we have mentioned
here thus seems to be $e^+ e^-$ collisions not far above the $\Upsilon(4S)$,
where the $\eta_C = +1$ and $\eta_C = -1$ states are not necessarily
equally populated.  Ultimately, however, the question is an experimental
one, and such effects can be studied at any energy and in any
configuration by searching for the $\sin \dmt$ term.

\section{Conclusions}

We have discussed the possibility of coherence of neutral $B$ meson pairs,
using a density-matrix approach which describes situations ranging from fully
correlated pairs to mixed (uncorrelated) states.  The density
matrix is parametrized by a ``polarization'' vector ${\bf Q'}$
describing a direction of ``quasi-spin.''  Usual experiments determine
only one component, $Q'_1$, of this vector, relating it to the dilution
factor in flavor tagging.  It gives rise to a characteristic modulation
of exponential decay by a $\cos \dmt$ term.  In general there can appear
a term proportional to $\sin \dmt$ as well, which has not been taken
into account in previous studies.  This term arises from the
component $Q'_2$ of the quasi-spin polarization vector, and is one
signal of coherence.  The component $Q'_3$ affects decays to CP
eigenstates, and can be searched for by studying such final states as
$J/\psi K_S$ and $J/\psi K_L$.  However, its investigation probably involves
{\it correlations} between decays to pairs of CP eigenstates, and thus may
require the production of a considerable number of $B$ mesons.
 
\section*{Acknowledgments}

We thank Yuval Grossman, John Jaros, Ady Mann,
Yoram Rozen, and Sheldon Stone for helpful discussions.
This work was supported in part by the United States Department of
Energy through Grant No.\ DE FG02 90ER40560 and by the U. S. -- Israel
Binational Science Foundation through Grant No.\  98-00237. J. L. R. wishes to
thank the Technion -- Israel Institute of Technology for gracious
hospitality during part of this work.

\def \ajp#1#2#3{Am.\ J. Phys.\ {\bf#1}, #2 (#3)}
\def \apny#1#2#3{Ann.\ Phys.\ (N.Y.) {\bf#1}, #2 (#3)}
\def \app#1#2#3{Acta Phys.\ Polonica {\bf#1}, #2 (#3)}
\def \arnps#1#2#3{Ann.\ Rev.\ Nucl.\ Part.\ Sci.\ {\bf#1}, #2 (#3)}
\def \art{and references therein}
\def \cmts#1#2#3{Comments on Nucl.\ Part.\ Phys.\ {\bf#1}, #2 (#3)}
\def \cn{Collaboration}
\def \cp89{{\it CP Violation,} edited by C. Jarlskog (World Scientific,
Singapore, 1989)}
\def \efi{Enrico Fermi Institute Report No.\ }
\def \epjc#1#2#3{Eur.\ Phys.\ J. C {\bf#1}, #2 (#3)}
\def \f79{{\it Proceedings of the 1979 International Symposium on Lepton and
Photon Interactions at High Energies,} Fermilab, August 23-29, 1979, ed. by
T. B. W. Kirk and H. D. I. Abarbanel (Fermi National Accelerator Laboratory,
Batavia, IL, 1979}
\def \hb87{{\it Proceeding of the 1987 International Symposium on Lepton and
Photon Interactions at High Energies,} Hamburg, 1987, ed. by W. Bartel
and R. R\"uckl (Nucl.\ Phys.\ B, Proc.\ Suppl., vol.\ 3) (North-Holland,
Amsterdam, 1988)}
\def \ib{{\it ibid.}~}
\def \ibj#1#2#3{~{\bf#1}, #2 (#3)}
\def \ichep72{{\it Proceedings of the XVI International Conference on High
Energy Physics}, Chicago and Batavia, Illinois, Sept. 6 -- 13, 1972,
edited by J. D. Jackson, A. Roberts, and R. Donaldson (Fermilab, Batavia,
IL, 1972)}
\def \ijmpa#1#2#3{Int.\ J.\ Mod.\ Phys.\ A {\bf#1}, #2 (#3)}
\def \ite{{\it et al.}}
\def \jhep#1#2#3{JHEP {\bf#1}, #2 (#3)}
\def \jpb#1#2#3{J.\ Phys.\ B {\bf#1}, #2 (#3)}
\def \lg{{\it Proceedings of the XIXth International Symposium on
Lepton and Photon Interactions,} Stanford, California, August 9--14 1999,
edited by J. Jaros and M. Peskin (World Scientific, Singapore, 2000)}
\def \lkl87{{\it Selected Topics in Electroweak Interactions} (Proceedings of
the Second Lake Louise Institute on New Frontiers in Particle Physics, 15 --
21 February, 1987), edited by J. M. Cameron \ite~(World Scientific, Singapore,
1987)}
\def \kdvs#1#2#3{{Kong.\ Danske Vid.\ Selsk., Matt-fys.\ Medd.} {\bf #1},
No.\ #2 (#3)}
\def \ky85{{\it Proceedings of the International Symposium on Lepton and
Photon Interactions at High Energy,} Kyoto, Aug.~19-24, 1985, edited by M.
Konuma and K. Takahashi (Kyoto Univ., Kyoto, 1985)}
\def \mpla#1#2#3{Mod.\ Phys.\ Lett.\ A {\bf#1}, #2 (#3)}
\def \nat#1#2#3{Nature {\bf#1}, #2 (#3)}
\def \nc#1#2#3{Nuovo Cim.\ {\bf#1}, #2 (#3)}
\def \nima#1#2#3{Nucl.\ Instr.\ Meth. A {\bf#1}, #2 (#3)}
\def \np#1#2#3{Nucl.\ Phys.\ {\bf#1}, #2 (#3)}
\def \PDG{Particle Data Group, D. E. Groom \ite, \epjc{15}{1}{2000}}
\def \pisma#1#2#3#4{Pis'ma Zh.\ Eksp.\ Teor.\ Fiz.\ {\bf#1}, #2 (#3) [JETP
Lett.\ {\bf#1}, #4 (#3)]}
\def \pl#1#2#3{Phys.\ Lett.\ {\bf#1}, #2 (#3)}
\def \pla#1#2#3{Phys.\ Lett.\ A {\bf#1}, #2 (#3)}
\def \plb#1#2#3{Phys.\ Lett.\ B {\bf#1}, #2 (#3)}
\def \pr#1#2#3{Phys.\ Rev.\ {\bf#1}, #2 (#3)}
\def \prc#1#2#3{Phys.\ Rev.\ C {\bf#1}, #2 (#3)}
\def \prd#1#2#3{Phys.\ Rev.\ D {\bf#1}, #2 (#3)}
\def \prl#1#2#3{Phys.\ Rev.\ Lett.\ {\bf#1}, #2 (#3)}
\def \prp#1#2#3{Phys.\ Rep.\ {\bf#1}, #2 (#3)}
\def \ptp#1#2#3{Prog.\ Theor.\ Phys.\ {\bf#1}, #2 (#3)}
\def \rmp#1#2#3{Rev.\ Mod.\ Phys.\ {\bf#1}, #2 (#3)}
\def \rp#1{~~~~~\ldots\ldots{\rm rp~}{#1}~~~~~}
\def \si90{25th International Conference on High Energy Physics, Singapore,
Aug. 2-8, 1990}
\def \slc87{{\it Proceedings of the Salt Lake City Meeting} (Division of
Particles and Fields, American Physical Society, Salt Lake City, Utah, 1987),
ed. by C. DeTar and J. S. Ball (World Scientific, Singapore, 1987)}
\def \slac89{{\it Proceedings of the XIVth International Symposium on
Lepton and Photon Interactions,} Stanford, California, 1989, edited by M.
Riordan (World Scientific, Singapore, 1990)}
\def \smass82{{\it Proceedings of the 1982 DPF Summer Study on Elementary
Particle Physics and Future Facilities}, Snowmass, Colorado, edited by R.
Donaldson, R. Gustafson, and F. Paige (World Scientific, Singapore, 1982)}
\def \smass90{{\it Research Directions for the Decade} (Proceedings of the
1990 Summer Study on High Energy Physics, June 25--July 13, Snowmass, Colorado),
edited by E. L. Berger (World Scientific, Singapore, 1992)}
\def \tasi{{\it Testing the Standard Model} (Proceedings of the 1990
Theoretical Advanced Study Institute in Elementary Particle Physics, Boulder,
Colorado, 3--27 June, 1990), edited by M. Cveti\v{c} and P. Langacker
(World Scientific, Singapore, 1991)}
\def \yaf#1#2#3#4{Yad.\ Fiz.\ {\bf#1}, #2 (#3) [Sov.\ J.\ Nucl.\ Phys.\
{\bf #1}, #4 (#3)]}
\def \zhetf#1#2#3#4#5#6{Zh.\ Eksp.\ Teor.\ Fiz.\ {\bf #1}, #2 (#3) [Sov.\
Phys.\ - JETP {\bf #4}, #5 (#6)]}
\def \zpc#1#2#3{Zeit.\ Phys.\ C {\bf#1}, #2 (#3)}
\def \zpd#1#2#3{Zeit.\ Phys.\ D {\bf#1}, #2 (#3)}


\begin{thebibliography}{99}
\bibitem{fact} Here we have inserted an overall factor of $e^{- \Gamma t}$
to correspond to the exponential decay of the state with direction $-\hp$.
It could just as well have been included in the (un-normalized)
density matrix as a common factor.
\end{thebibliography}
\end{document}